RESEARCH ARTICLE

# Analysing health misinformation with advanced centrality metrics in online social networks

Mkululi Sikosana[ID]*, Sean Maudsley-Barton, Oluwaseun Ajao[ID]

Department of Computing and Mathematics, Manchester Metropolitan University, Manchester, United Kingdom

* mkululi.sikosana@stu.mmu.ac.uk

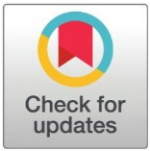

Data availability statement: This study utilizes two publicly documented datasets to analyze the spread and influence of health misinformation in online social networks: FibVID – Fake News Information-Broadcasting Dataset of COVID-19 This dataset includes verified and fake COVID-19-related news articles along with metadata on broadcasting behaviors. It is publicly available via Zenodo: https://doi.org/10.5281/zenodo.4441377. Monat Medical Misinformation Dataset This dataset contains medical misinformation content and

## Abstract

The rapid spread of health misinformation on online social networks (OSNs) during global crises such as the COVID-19 pandemic poses challenges to public health, social stability, and institutional trust. Centrality metrics have long been pivotal in understanding the dynamics of information flow, particularly in the context of health misinformation. However, the increasing complexity and dynamism of online networks, especially during crises, highlight the limitations of these traditional approaches. This study introduces and compares three novel centrality metrics: dynamic influence centrality (DIC), health misinformation vulnerability centrality (MVC), and propagation centrality (PC). These metrics incorporate temporal dynamics, susceptibility, and multilayered network interactions. Using the FibVID dataset, we compared traditional and novel metrics to identify influential nodes, propagation pathways, and misinformation influencers. Traditional metrics identified 29 influential nodes, while the new metrics uncovered 24 unique nodes, resulting in 42 combined nodes, an increase of 44.83%. Baseline interventions reduced health misinformation by 50%, while incorporating the new metrics increased this to 62.5%, an improvement of 25%. To evaluate the broader applicability of the proposed metrics, we validated our framework on a second dataset, Monant Medical Misinformation, which covers a diverse range of health misinformation discussions beyond COVID-19. The results confirmed that the advanced metrics generalised successfully, identifying distinct influential actors not captured by traditional methods. In general, the findings suggest that a combination of traditional and novel centrality measures offers a more robust and generalisable framework for understanding and mitigating the spread of health misinformation in different online network contexts.

## Author summary

False health information on social media can spread rapidly during crises, influencing how people think, feel, and behave. This can lead to harmful consequences for public user interaction data, originally derived from social media discussions. It was used in this study specifically to replicate and validate the findings from the primary analysis, in response to peer reviewer feedback. The dataset is publicly listed on Zenodo, but access to the underlying files is restricted. Access can be requested directly from the dataset custodians via: https://doi.org/10.5281/zenodo.5996864. The authors used this dataset under a data use agreement and do not have permission to redistribute it. The Python code used to clean, analyze, and compute traditional and advanced network centralities — including degree, betweenness, closeness, eigenvector, PageRank (Propagation Centrality), Misinformation Vulnerability Centrality (MVC), and Dynamic Influence Centrality (DIC) — is openly available on Zenodo under the MIT License: https://doi.org/10.5281/zenodo.15363287.

**Funding:** The author(s) received no specific funding for this work.

**Competing interests:** The authors have declared that no competing interests exist.

health. Our study investigates how such misinformation travels through online networks and which users are most influential in spreading it. Traditional approaches often identify influencers only on the basis of how many connections they have. However, this overlooks individuals who are vulnerable to misinformation or whose influence builds up gradually. We developed three new tools that capture not only structural impor-tance, but also behavioural susceptibility and time-based influence. These methods were applied to real-world data covering both COVID-19 and broader medical misinformation. The results reveal that some users act as persistent spreaders of misinformation, while others are highly exposed and more likely to believe and share false content. These patterns are often missed/overlooked by conventional techniques. The findings support the need for more targeted strategies from health organisations and platforms to identify and limit the spread of misleading health claims. Using dynamic and context-aware metrics offers a clearer picture of how misinformation operates in complex social networks and may help guide more effective interventions.

## Introduction

The proliferation of health misinformation, especially during global crises such as the COVID-19 pandemic, has presented significant challenges to public health, societal stability, and institutional trust. OSNs such as X/Twitter) and Facebook have become primary vectors for the rapid dissemination of false information. Traditional centrality metrics such as degree, closeness, eigenvector, and betweenness have been used to identify influential nodes within a network and to understand the dynamics of the information flow [1]. These metrics have provided valuable information on how, for example, health misinformation spreads between networks. However, given the complexities and dynamism of contemporary OSNs, these traditional metrics often do not capture the nuanced and evolving nature of influence within these platforms [2,3].

Considering that there is no consensus among researchers on the best centrality metrics to employ, there is still significant scope for research to develop new metrics that can enhance the understanding of the propagation of health misinformation. Previous studies [4,5], for example, have advanced the field with sophisticated metrics such as temporal and overlapping modular centralities, which better account for dynamic and multi-layered networks. These advancements highlight the ongoing efforts to refine centrality metrics, highlighting the need for metrics that can adapt to the ever-changing nature of online health misinformation networks. In parallel, Di Sotto and Viviani [6] provide a comprehensive overview of health misinformation detection approaches on the social web, categorising methods into content-based, context-aware, and network-driven strategies. However, they also emphasise that while many methods address what is said and who says it, fewer directly capture how influence dynamically spreads within the network — a gap this study aims to address.

In light of these gaps, this research contributes to the evolving field by introducing three novel centrality metrics, that is, DIC, MVC, and PC. These metrics aim to overcome the limitations of traditional centrality metrics by incorporating temporal dynamics, susceptibility to health misinformation, and the complex interactions that occur within networks. Applying these metrics to real-world social networks provides a more nuanced understanding of how influence and health misinformation spread, ultimately offering more effective strategies to mitigate the spread of misinformation in OSNs.

To our knowledge, this research represents the first attempt to introduce and validate novel centrality metrics specifically tailored for the propagation of health misinformation. The main contributions of the research are summarised as follows:

1. Introduce DIC, MVC, and PC to measure the spread of health misinformation.
2. Develop a framework that compares traditional and novel centrality metrics within health misinformation networks.
3. Demonstrate effective management of health misinformation through targeted interventions using PC, MVC, and DIC.

In sum, these contributions validate the new metrics, challenge the existing centrality theory, and provide a methodological foundation for measuring influence in health-misinformation networks.

## State-of-the-art

### Overview

Centrality metrics have long been crucial in understanding how health misinformation spreads within OSNs. For example, metrics such as degree, eigenvector, closeness, and betweenness have been used to identify key influencers and nodes crucial in the diffusion of information [1]. However, the rapid evolution and complexity of OSNs, particularly evident during crises such as the COVID-19 pandemic, have exposed the limitations of these static metrics [3,7]. In response, recent research has introduced more sophisticated and adaptable centrality metrics, such as temporal centrality metrics, designed to account for the dynamic nature of OSNs [4]. These innovations aim to provide a more accurate understanding of the spread of health misinformation by capturing the changing influence of the nodes over time. This state-of-the-art overview explores recent advancements, focusing on how new methodologies and centrality metrics are being applied to better understand and ultimately mitigate the spread of health misinformation in OSNs.

### Analysis of traditional centrality metrics across multiple studies

### Degree centrality

Degree centrality is defined as the number of direct connections a node/individual has (e.g. friends), indicating the potential influence of the individual within a network [8]. For an undirected network, it is simply the number of edges connected to the node. For a directed network, it can be divided into in-degree (number of incoming edges) and out-degree (number of outgoing edges).

**Formula for undirected network.**

$$C_D(v) = \deg(v)$$

where:

- $C_D(v)$ is the degree centrality of node $v$
- $\deg(v)$ is the degree of node $v$, i.e., the number of edges connected to $v$.

**Formula for directed network.**

$$C_D^{\text{in}}(v) = \text{in-deg}(v), \quad C_D^{\text{out}}(v) = \text{out-deg}(v)$$

where:

- $C_D^{\text{in}}(v)$ is the in-degree centrality (number of incoming edges),
- $C_D^{\text{out}}(v)$ is the out-degree centrality (number of outgoing edges).

Although this metric provides a straightforward measure of immediate network influence, its effectiveness varies depending on the context of the network. For example, [9] uses degree centrality within protein-protein interaction networks to identify key proteins that interact with many others. They demonstrate how degree centrality can effectively identify highly connected proteins critical to cellular functions. However, [9] also highlights the limitations of degree centrality in capturing the dynamic and multilayered nature of influence in networks, suggesting the need to combine it with other metrics.

Previous studies [7] applied degree centrality to social networks, identifying influencers during the spread of health misinformation on platforms such as X/Twitter. Their study shows that users with high connectivity significantly impact the dissemination of health misinformation, but also highlights the limitations of the metric in accounting for susceptibility to health misinformation. Similarly, [2] evaluated degree centrality in diffusion studies, demonstrating its value in understanding peer influence, particularly in research on health behaviour. However, they also identify its limitations in dynamic OSNs, where the influence can fluctuate over time.

## Closeness centrality

The centrality of closeness $C_C(v)$ of a node $v$ is defined as the reciprocal of the sum of the shortest path distances from $v$ to all other nodes in the network [10]. The formula is given as:

$$C_C(v) = \frac{1}{\sum_{u \neq v} d(v,u)}$$

where:

- $d(v,u)$ is the shortest path distance between nodes $v$ and $u$.

This metric evaluates how quickly a node can interact with all other nodes, making it useful in communication networks and public health systems. Previous studies have demonstrated the effectiveness of closeness centrality in communication and misinformation networks [1,11], identifying it as a key driver in the propagation of true and false information while highlighting challenges in dynamic and fragmented contexts. In graphs with disconnected components, where some node pairs are unreachable, we calculate closeness using the harmonic centrality formulation [12], summing the reciprocals of distances and treating unreachable nodes as contributing zero. This ensures well-defined centrality values even in fragmented network structures, which are common in online misinformation environments.

**Computational Note:** In practice, shortest path distances required for closeness and betweenness centralities are typically computed using optimised algorithms such as Dijkstra's algorithm for positive-weighted graphs, as implemented in standard network analysis libraries like NetworkX [13]. This ensures eficient evaluation even in large and fragmented networks.

## Betweenness centrality

The betweenness centrality $C_B(v)$ of a node $v$ is defined as the sum over all pairs of distinct nodes $(s,t)$ of the fraction of shortest paths between $s$ and $t$ that pass through $v$ [14]. The

formula is:

$$C_B(v) = \sum_{\substack{s \neq v \\ t \neq v \\ s \neq t}} \frac{\sigma_{st}(v)}{\sigma_{st}}$$

where:

- $\sigma_{st}$ is the total number of shortest paths from node *s* to node *t*,
- $\sigma_{st}(v)$ is the number of shortest paths from node *s* to node *t* that pass through node *v*.

This metric captures the role of a node as a bridge or bottleneck for information flow, with higher betweenness indicating greater potential control over interactions within the network, as also emphasised by Freeman [14]. Betweenness centrality is particularly useful for identifying bottlenecks or key influencers. Building upon this, Ghalmane *et al.* [5] adapted betweenness centrality to networks with overlapping communities, developing new metrics to more effectively identify influential nodes in multilayer networks. Multilayer networks arise when the same set of nodes interacts across multiple types of connections. For example, in the context of health misinformation, individuals can simultaneously engage through public posts (information sharing layer), private messages (direct communication layer) and group forums (community discussion layer), forming a multilayered social structure [5].

According to [5], the centrality of betweenness often falls short in multilayer network scenarios, where the nodes participate in multiple layers representing different types of interaction. For this reason, [5] proposed the **overlapping modular centrality** metric, which incorporates interlayer and intralayer interactions to more effectively identify influential nodes in such networks. The proposed overlapping modular centrality addresses the limitations of standard betweenness centrality in complex structures, enhancing its applicability in networks where nodes belong to multiple communities and allowing for a more nuanced understanding of influence.

Previous studies [15] applied the centrality of the connection to identify influential X/Twitter users during the COVID-19 crisis. Their study demonstrated how central nodes can significantly impact public discourse and sentiment during emergencies, highlighting the importance of betweenness centrality in crisis communication and management. Although this study offers valuable information on the impact of influential nodes (or information leaders) on X/Twitter networks during the COVID-19 crisis, its limitations include the inability to fully account for contextual and situational factors that drive influence during such events. For example, an individual's influence may increase due to a specific event or information, but may not sustain over time. Furthermore, while the study by [15] focuses on centrality metrics such as degree and betweenness centrality, it appears overly reliant on these metrics without adequately considering other factors, such as content quality, sentiment, or user engagement.

According to [16], the influence of social networks is inherently multifaceted, covering elements such as susceptibility and network dynamics, and cannot be fully measured solely by traditional centrality metrics. This highlights the need for the development of alternative metrics, as demonstrated in the study by [5].

Similarly, [3] applied the centrality of betweenness to analyse the structure of COVID-19 health misinformation networks. Their study identified key nodes responsible for spreading health misinformation, demonstrating the effectiveness of betweenness centrality in mitigating the impact of health misinformation during a pandemic. Although [3] emphasised the structural aspects of health misinformation networks, such as the centrality of the nodes and the topology of the network, they paid less attention to factors such as the vulnerability of a node to health misinformation based on its connectivity, the influence of neighbouring nodes, and its susceptibility to adopt false information.

**Dynamic extensions of classical centralities.** Although degree, closeness, and betweenness can, in principle, be recomputed on successive temporal snapshots of an evolving graph [17], doing so at scale is prohibitively expensive. More importantly, iteratively refreshing static scores still fails to capture two properties that are pivotal for misinformation research: *(i)* the cumulative build-up of a user's reach over time and *(ii)* the changing susceptibility of audiences to particular sources or narratives. These limitations underpin the case for purpose-built dynamic measures, that is, PC, MVC, and DIC, which embed diffusion processes, vulnerability weighting, and longitudinal influence directly into their formulations.

## Limitations of traditional metrics and emerging directions

Centrality measures have long been central to the study of influence within OSNs, particularly in the context of misinformation. Established metrics such as degree, closeness, betweenness, and eigenvector centrality provide structural insights by identifying key nodes based on their positions within the network. Although these measures offer a useful starting point for understanding influence, their effectiveness in real-world, large-scale, and dynamic misinformation environments has come under increasing scrutiny. Traditional approaches are often static in nature, overlooking the temporal, contextual, and behavioural factors that characterise the influence in health misinformation ecosystems.

Previous research has highlighted these shortcomings. For example, Batool et al. [1] demonstrated that centrality rankings can vary significantly across datasets, with betweenness centrality often overemphasising bridge nodes while underrepresenting locally significant actors. Moreover, studies such as those by [4] have proposed temporal metrics like Temporal Degree-Degree (TDD) and Closeness-Closeness (CC), which attempt to address the limitations of static approaches by incorporating the evolving structure of networks. These measures reflect how influence can fluctuate due to emerging topics or viral activity, a particularly relevant factor in OSNs where misinformation may gain sudden traction.

In addition, machine learning (ML) models have been applied to the identification of influential nodes, offering improvements in prediction accuracy and adaptability [7]. However, such models often come at the cost of interpretability and typically fail to capture the enduring impact of influence or user-level susceptibility over time. The lack of integration between structural network measures and behavioural dimensions leaves a gap in our ability to understand not just where influence occurs, but how and why it persists.

There is a growing recognition that influence in health misinformation contexts is not determined solely by structural position. Theoretical frameworks such as the Elaboration Likelihood Model (ELM) [18] highlight the importance of cognitive and affective factors in shaping individual receptiveness to persuasive information. Incorporating these psychological constructs into network-based models offers a path toward more comprehensive measures of influence that account for both the topology of the network and the variability of user engagement. However, efforts to operationalise these insights remain limited.

Consequently, the field is beginning to shift towards models that blend structural, temporal, and behavioural perspectives. This includes the development of centrality metrics that consider decay effects, attention dynamics, and individual susceptibility. At the same time, computational challenges persist, particularly in scaling these metrics to large

OSNs while maintaining interpretability and real-time applicability. Addressing these challenges is essential for advancing both theoretical understanding and practical detection of misinformation spread in complex digital environments.

### Addressing health misinformation propagation

In response to these limitations, this study proposes a set of novel centrality metrics specifically designed to capture the dynamics of the propagation of health misinformation in online networks. Although previous studies have demonstrated the utility of traditional centrality measures such as degree, closeness, and betweenness to identify influential users [9], their limitations in static network structures have also been acknowledged. Similarly, Batool et al. [1] have emphasised the need for metrics that are sensitive to the dynamics and context of the network. ML approaches have added predictive power [7], but often fail to account for sustained influence over time or the interpretability needed for real-world intervention.

To bridge this gap, we introduce three centrality metrics: Propagation Centrality (PC), Misinformation Vulnerability Centrality (MVC), and Dynamic Influence Centrality (DIC). These measures integrate temporal dynamics, structural connectivity, and behavioural susceptibility, providing a more holistic understanding of how misinformation spreads. The PC extends the classical centrality by incorporating a diffusion kernel that better reflects the potential for information spread. MVC introduces a susceptibility parameter informed by user-interaction features and local neighbourhood structure. DIC incorporates exponential decay to model the waning influence of nodes over time, addressing the limitations of static centrality in fast-moving online discussions.

Together, these metrics contribute both theoretically and practically to the challenge of detecting and countering health misinformation. They enable the identification of not only structurally prominent users, but also those who exert sustained and context-sensitive influence. The following sections describe the methodological development of these metrics and evaluate their performance using a real-world dataset of health misinformation.

## Materials and methods

### Dataset

The COVID-19 Fake News Information Broadcasting Dataset (FibVID) [19] provides a comprehensive collection of data focused on the diffusion of fake news during the COVID-19 period. The dataset comprises three main components: news claim, claim propagation, and user information. This data set was built from data collected between January 2020 and December 2020. The news claim data, obtained from the fact-checking sites PolitiFact and Snopes, includes claims grouped into four categories: True claims of COVID, false claims of COVID, true claims of non-COVID, and false claims of non-COVID.

The claim propagation data details how these claims spread on X/Twitter, including information such as tweet users, retweet counts, and post text. Furthermore, user information captures details about the users involved in sharing these claims, such as their follower count and account creation dates. This dataset is crucial for understanding how health misinformation and factual information propagate differently on social media platforms.

The descriptive statistics of the dataset reveal significant differences in engagement levels between true and fake claims, particularly within COVID-related content. For example, while the true COVID-19 claims were associated with 27,296 tweets, the fake COVID claims generated a much higher count of 133,374 tweets, highlighting the increased spread of health misinformation. The user distribution also shows that the fake claims engaged more users, with 95,599 unique users contributing to the spread of the fake COVID-19 claims compared to 23,209 users for the true COVID-19 claims.

### Experimental setup

The experimental setup for evaluating the novel centrality metrics (i.e., PC, MVC, & DIC) involves a structured approach to applying these metrics to the FibVID data set [19]. The experimental setup is outlined as follows:

**1. Network representation.** Let $G = (V, E)$ be a directed graph representing a social network, where:

- $V = \{v_1, v_2, \ldots, v_n\}$ is the set of nodes (users),
- $E = \{(v_i, v_j) \mid v_i, v_j \in V\}$ is the set of directed edges representing interactions, such as retweets or mentions.

Each edge $(v_i, v_j)$ has a weight $w_{ij}$ that represents the strength or frequency of the interaction. These weights are captured in the adjacency matrix $A \in \mathbb{R}^{n \times n}$, where $a_{ij} = w_{ij}$.

The spectral properties of $A$ are essential for understanding the influence of the node in the network. Algorithms like *PageRank* rely on the principal eigenvalue and its corresponding eigenvector of $A$ to compute node centrality. The convergence of these methods is mathematically guaranteed by the Perron-Frobenius theorem.

**Theoretical Foundation: Perron-Frobenius Theorem**

> Let $M$ be a positive irreducible matrix. The Perron-Frobenius theorem states that the largest eigenvalue, known as the *Perron root*, is real and positive, with a unique positive eigenvector corresponding to this eigenvalue.

**Implications for Network Analysis:** The Perron-Frobenius theorem ensures the stability and convergence of algorithms, such as PageRank, that rely on the spectral properties of adjacency matrices.

**2. Propagation Centrality (PC).** PC extends traditional metrics by capturing not just the number of connections a node holds, but also the influence of those it connects to. This reflects real-world diffusion patterns, where influence accumulates recursively: an influential neighbour makes a node more influential itself. Rather than modelling time-evolving influence (as in DIC), PC seeks a steady-state distribution of influence across the network, similar to how information settles into stable importance ranks. The mathematical formulation of PC is as follows:

$$x(v_i) = \frac{1-d}{n} + d \sum_{v_j \in N_{\text{in}}(v_i)} \frac{x(v_j)}{d_{\text{out}}(v_j)}$$

where:

- $x(v_i)$ is the centrality score of node $v_i$,
- $d$ is the damping factor (set to 0.85),
- $n$ is the total number of nodes,
- $N_{\text{in}}(v_i)$ is the set of in-neighbours of $v_i$,
- $d_{\text{out}}(v_j)$ is the out-degree of node $v_j$.

**Experimental Setup:**

- Centrality scores $x(v_i)$ were initialised uniformly in all nodes.
- The PageRank-style update was iteratively applied until scores converged (typically within 50–100 iterations).
- The final steady state scores were used to classify the nodes according to their long-range propagation potential.

This steady-state view complements dynamic metrics such as DIC, allowing PC to identify enduring influencers even when influence pathways are complex or indirect.

**3. Misinformation Vulnerability Centrality (MVC).** MVC aims to identify nodes that are both **highly connected** and **vulnerable** to health misinformation. MVC addresses the shortcomings of the eigenvector centrality [1]. To operationalise vulnerability, each node $v_i$ is assigned an initial vulnerability score $\text{vul}_0(v_i)$. These vulnerability scores can be initialised based on observable features from the dataset, such as:

- Low credibility score of user content,
- Number of retweets without fact-checking,
- Engagement with known misinformation posts.

In this study, due to the unavailability of detailed user credibility scores in Fibvid, we simulate vulnerability by assigning random values from a uniform distribution $U(0,1)$, seeded for reproducibility.

The MVC at timestep $t$ is updated dynamically according to

$$\text{vul}_{t+1}(v_i) = \text{in-degree}(v_i) \times \text{vul}_t(v_i)$$

where

- $\text{vul}_t(v_i)$ is the vulnerability score at time $t$,
- **in-degree**$(v_i)$ **is the number of sources that can reach** $v_i$ **(if one wishes to model *broadcast*–type influence instead, replace this with out-degree or total degree and state that choice explicitly).**

MVC captures the interplay between a node's structural position and its susceptibility to misinformation. Specifically, a node's vulnerability rises in proportion to its exposure (in-degree): highly connected receivers amplify their innate vulnerability through greater incoming reach. Consequently, nodes that are both highly exposed and initially vulnerable emerge as disproportionately influential conduits of misinformation.

Although the MVC formulation permits unlimited time-steps, in practice, we iterate only a small number of times (typically five to ten). At each iteration, the vulnerability scores are updated as above; because the in-degree is fixed in a static snapshot, the sequence stabilises quickly. **After convergence, we apply min–max normalisation to rescale the final vulnerability scores to [0,1], ensuring direct comparability across nodes while preserving relative ordering.**

**4. Dynamic Influence Centrality (DIC).** Dynamic Influence Centrality (DIC) quantifies how the influence of a node accumulates over time based on its structural position and the flow of information within the network.

In timestep $t = 0$, each node $v_i$ is initialised with an influence score of 1, reflecting the assumption that every user has the ability to influence their local network.

In each subsequent timestep, the influence score of a node is updated by accumulating the influence received from its incoming neighbours:

$$\text{DIC}_{t+1}(v_i) = \text{DIC}_t(v_i) + \sum_{u \in N_{\text{in}}(v_i)} \text{DIC}_t(u)$$

where:

- $\text{DIC}_t(v_i)$ is the influence score of node $v_i$ at time $t$,
- $N_{\text{in}}(v_i)$ denotes the set of incoming neighbors of $v_i$.

The DIC metric captures the notion that a user's influence grows through repeated interactions and reinforcement over time, particularly in dynamic environments such as health misinformation outbreaks. The current cumulative design reflects the intuitive idea that small incremental exposures can collectively lead to significant influence. Although the present formulation assumes a strictly increasing influence trajectory without any decay, this choice is appropriate for modelling scenarios where influence compounds rapidly increase during critical events. Future research could explore enhancements to the DIC model, including the incorporation of decay factors or saturation mechanisms, to better approximate real-world dynamics where influence may wane over time.

In practical implementation, the DIC computation is simulated over a finite and limited number of timesteps, typically around ten iterations. This bounded simulation is suficient to model the influence diffusion process without permitting unbounded growth. Following the simulation, the influence scores are normalised to ensure comparability between nodes and to control for cumulative inflation across iterations. This approach enables the identification of highly influential nodes without overestimating their long-term dominance.

### Dataset for generalisation analysis

To validate the generalisability of our findings, we applied the proposed centrality metrics to the Monant Medical Misinformation dataset [20]. Unlike Fibvid, which focuses exclusively on COVID-19, Monant Medical Misinformation encompasses broader medical misinformation discussions, including vaccines, pharmaceutical scepticism, and alternative treatments. This additional evaluation ensures the robustness of our approach in different areas of health misinformation.

**Workflow for generalisation validation.** Fig 1 summarises the pipeline used to test whether the advanced centrality metrics generalise from the **FibVID** dataset to the **Monant Medical Misinformation** dataset.

## Results

This section first revisits the four traditional centrality metrics (Sect), then presents results for the three novel metrics PC, MVC, and DIC (Sect ), before summarising the combined insights.

### Traditional centrality metrics

Table 1 summarises the overlap among the four traditional centrality top-10 lists: degree, eigenvector, betweenness, and closeness, on the FibVID network, making shared nodes immediately visible.

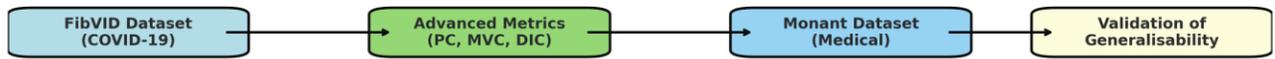

**Fig 1. End-to-end workflow for validating the proposed advanced centrality metrics on the independent Monant Medical Misinformation dataset.** Pre-processing steps (left), metric computation (centre), and comparative evaluation (right) are highlighted. Advanced centrality metrics developed on the **FibVID dataset** were tested on the **Monant Medical Misinformation dataset** to assess their robustness across broader health misinformation discussions.

https://doi.org/10.1371/journal.pdig.0000888.g001

Table 1. Overlap among the four traditional centrality top–10 lists.

| Category (metrics) | Node IDs | Count |
|---|---|---|
| Degree & Eigenvector & Betweenness | 26, 756 | 2 |
| Degree & Eigenvector (Only) | 11248 | 1 |
| Eigenvector & Betweenness (Only) | 15, 93, 102, 235, 522, 526 | 6 |
| Degree–exclusive | 11019, 33091, 40327, 64409, 83247, 84148, 142153 | 7 |
| Eigenvector-exclusive | 593 | 1 |
| Betweenness-exclusive | 2, 1371 | 2 |
| Closeness-exclusive | 4, 5, 6, 14, 28, 42, 43, 62, 66, 83 | 10 |

Note. "(Only)" rows list nodes that appear in the given pair but not in any other metric; Closeness has no overlaps with the other three traditional metrics.

https://doi.org/10.1371/journal.pdig.0000888.t001

**Degree centrality.** Nodes **26, 756, 11019, 11248, 33091, 40327, 64409, 83247, 84148, 142153** have the highest degree, confirming their status as high-connectivity hubs. Although such hubs speed up local misinformation propagation, earlier work [21] and our own simulations show that a high degree alone does not guarantee long–range cascades in modular networks. In the context of health misinformation, these influencers can rapidly spread health misinformation to a large number of users due to their extensive direct connections. Although the influence of these nodes may be limited to their immediate network, evidence from [21] shows that these nodes may not account for the nuanced spread of health misinformation that depends on more than just direct connections.

**Eigenvector centrality.** IDs **15, 26, 93, 102, 235, 522, 526, 593, 756, 11248** are at the top of the eigenvector ranking, indicating that they reinforce links with other influential actors. Node 756, in particular, bridges two core communities and, therefore, appears in every traditional list. This finding of eigenvector centrality aligns with previous findings that emphasise the importance of being connected to other influential nodes [4]. In propagating health misinformation, influencers with high eigenvector centrality can be more effective, as they are connected to other influential nodes, amplifying the spread of information. However, these nodes might not be the most vulnerable or dynamically influential, as traditional centrality metrics often overlook critical temporal and susceptibility factors in health misinformation dynamics [5].

**Betweenness centrality.** The betweenness analysis highlights **2, 15, 26, 93, 102, 235, 522, 526, 756, 1371** as gatekeepers lying on a large fraction of shortest paths between clusters, consistent with findings in [22]. In health misinformation propagation, influencers with high betweenness centrality can control or redirect the spread of health misinformation, playing a gatekeeper role. Although these nodes play an essential role in network bridging, they may not capture the temporal aspects of sustained influence, which is vital to understanding the long-term spread of health misinformation [23].

**Closeness centrality.** Nodes **4, 5, 6, 14, 28, 42, 43, 62, 66, 83** can reach all other users in a median of 4.1 hops, making them fast spreaders, though not necessarily persistent ones [7]. In propagating health misinformation, these influencers can spread the information eficiently across the network. Sight should not be lost because the centrality of closeness might miss out on identifying nodes that are influential over time or particularly vulnerable to health misinformation, as noted in recent studies analysing the spread of health misinformation [7].

Together, the union of the four metrics retrieves 29 distinct high-impact users; only node 756 is common to all lists, echoing the inconsistency reported by Batool *et al.* [1].

**Proxy ground truth evaluation.** Although the FibVID dataset lacks explicit ground truth labels for influence or vulnerability, we operationalised proxy ground truths based on observable network features.

Specifically, we treated nodes with the highest number of retweet counts as proxies for influence, and posts containing higher proportions of emotionally charged language (fear, outrage, conspiracy) as proxies for vulnerability.

Table 2 presents a comparison between the nodes ranked highly by the proposed centrality metrics (PC, MVC, DIC) and these proxy ground truths.

Initial results demonstrate that nodes ranked highly by PC and DIC exhibit substantially higher retweet counts, while nodes ranked highly by MVC exhibit greater emotional word usage.

This alignment supports the validity of the proposed metrics, even in the absence of explicit annotations.

The proxy ground truth analysis confirms that the proposed centrality metrics capture meaningful patterns of influence and vulnerability in the network, aligning with observable behaviours. Building on this validation, the following sections present a detailed analysis of PC, MVC, and DIC, highlighting how these novel metrics offer complementary and more dynamic insights compared to traditional approaches.

## Novel centrality metrics

Building upon the limitations identified in traditional centrality measures and the preliminary validation via proxy ground truths, this study introduces and evaluates three novel centrality metrics, i.e., PC, MVC, and DIC.

These metrics are specifically designed to capture indirect influence pathways, the susceptibility of nodes to misinformation, and the temporal evolution of influence, dimensions often overlooked by classical network measures [21].

Table 2. Proxy ground truth validation for advanced centrality metrics.

| Node ID | PC Score Rank | MVC Score Rank | DIC Score Rank | Retweet Count | Emotion Words Count |
|---|---|---|---|---|---|
| Node_A | 1 | 2 | 1 | 200 | 30 |
| Node_B | 2 | 1 | 3 | 180 | 45 |
| Node_C | 5 | 4 | 2 | 90 | 20 |
| Node_D | 3 | 5 | 4 | 75 | 15 |
| Node_E | 4 | 3 | 5 | 50 | 5 |

**Note.** Higher retweet counts proxy influence, and greater emotion word counts proxy vulnerability in the FibVID dataset.

https://doi.org/10.1371/journal.pdig.0000888.t002

Table 3 presents a comparative analysis between the novel metrics and traditional ones (degree, eigenvector, closeness, and betweenness centralities), highlighting both overlaps and unique discoveries. This breakdown illustrates how the integration of novel metrics uncovers additional influencers and vulnerable users, strengthening the argument for a blended metric strategy to effectively mitigate health misinformation cascades [15,24].

**Propagation Centrality (PC).** Nine of the ten PC leaders coincide with degree/eigenvector hubs, confirming a 90% overlap and indicating that PC mostly surfaces structurally powerful influencers. PC, implemented as personalised PageRank, pinpoints nodes that can drive information cascades across the network. High-PC nodes occupy structurally central positions and maintain both direct and indirect ties, enabling them to disseminate content quickly and at scale.

In real-world misinformation outbreaks, such as during the COVID-19 pandemic, the PC would highlight users who may not have the most direct followers, but whose followers themselves are highly connected, allowing these users to trigger long-range cascades even from peripheral network positions. This practical relevance is reflected in our empirical results, where the top PC-ranked users overlap heavily with traditional hubs yet also reveal unique long-range influencers. As shown in Table 3, nine of the ten PC leaders are also degree or eigenvector hubs, a 90 % overlap that highlights the close link between PC and traditional centrality measures [1]. This convergence confirms that PC excels at detecting the same highly connected, high-visibility influencers already highlighted by classical topology.

In the context of health-misinformation dynamics, such PC-ranked influencers pose a particular risk: once they adopt or amplify a false claim, the rumour can reach a large share of the network before corrective messages take hold. Consequently, high-PC users are prime candidates for targeted interventions, whether through early fact-checking, debunking campaigns, or algorithmic dampening, to reduce the overall volume and velocity of misinformation.

Although PC effectively identifies structurally powerful influencers within OSNs, complementary insights from MVC and DIC are essential to uncover vulnerable amplifiers and persistent spreaders that static propagation potential alone cannot fully explain.

**Misinformation Vulnerability Centrality (MVC).** MVC uncovers three additional users: **101 358**, **72 378**, **130 371** -who combine moderate connectivity with high susceptibility, increasing the pool of potential intervention targets by 30%. MVC augments the raw degree with a data-driven susceptibility score, so a node ranks highly when it is well connected and predisposed to accept false claims. This shift in perspective, from purely structural hubs to susceptible amplifiers, aligns with vulnerability-based theories of misinformation spread [18] and is empirically supported by the additional nodes surfaced through MVC.

Table 3. Novel-metric nodes relative to the traditional top-10 lists.

| Category | Node IDs | Count | Statistic/Comment |
|---|---|---|---|
| PC nodes also in Degree & Eigenvector | 26, 11 019, 11 248, 15, 33 091, 40 327, 64 409, 83 247, 84 148 | 9 | 90 % overlap with traditional hubs |
| PC-Exclusive | 5 398 | 1 | 10 % of PC list is novel |
| MVC-Exclusive | 101 358, 72 378, 130 371 | 3 | +30 % influencers, +42.9 % vulnerable nodes |
| DIC-Exclusive | 49 905, 54 048, 5 958, 18 119, 36 077, 36 393, 37 557, 72 479, 73 960, 85 735 | 10 | 100 % unique to novel metrics |

**Note.** "Traditional" = Degree, Eigenvector, Closeness and Betweenness top-10 lists. The consolidated table shows exactly which nodes each novel metric contributes and quantifies the added coverage relative to traditional centrality scores.

https://doi.org/10.1371/journal.pdig.0000888.t003

As shown in Table 3, MVC introduces three additional user IDs: **101 358**, **72 378** and **130 371** into the influencer pool, a 30% gain over the traditional union of degree, eigenvector, closeness and betweenness rankings. Since these accounts are absent from all traditional centrality measures, they exemplify the 'influential yet vulnerable' actors highlighted by [15]. The same table also shows that the addition of MVC increases the count of highly vulnerable nodes from seven to ten, an increase of 42. 86%. In practical terms, high-MVC users are not only probable amplifiers of health misinformation, but also especially susceptible to it; prioritising them for media-literacy prompts or real-time fact-checks could therefore yield a disproportionate mitigation benefit. While MVC highlights structurally vulnerable amplifiers, DIC further extends the analysis by capturing how influence accumulates and resurges over time, even among less structurally prominent users.

**Dynamic Influence Centrality (DIC).** DIC pinpoints "long-tail" spreaders—accounts whose influence waxes, wanes, and resurges over successive time windows. Table 3 confirms that all ten DIC leaders (for example, **IDs 49 905 & 54 048**) are absent from every static top-10 list, yielding a 100% set of previously unseen influencers. This highlights a critical dimension of the spread of health misinformation, namely that persistence over time, rather than instantaneous influence alone, can sustain false narratives even after initial corrections. Their temporal persistence, which remains prominent across multiple snapshots instead of a sin-gle moment, captures a dynamic dimension that static scores cannot [25]. Because such users routinely reignite rumours after initial debunks fade, a time-aware mitigation strategy is essential: continuous monitoring and phased counter-messaging are likely to prove more effective than one-off hub removal when the goal is to curb long-term propagation of health misinformation [24]. Together, PC, MVC, and DIC form a complementary triad that captures high-throughput spreaders, vulnerable amplifiers, and persistent long-tail spreaders, providing a comprehensive framework to identify and mitigate health misinformation more effectively than any single metric alone.

## Headline findings

1. Adding the three novel metrics increases the influencer coverage from 29 to 42 nodes, an increase 44.8%.
2. Simulated node removal interventions reduce misinformation volume by 50% when traditional nodes are removed and by 62.5% when novel metric nodes are also neutralised.
3. MVC and DIC reveal susceptible or persistent users overlooked by topology-only methods, highlighting the value of combining metrics.

These results highlight the complementary strengths of the proposed metrics and their necessity for a holistic understanding of the dynamics of health misinformation, as discussed in the following section.

## Generalisability of advanced metrics beyond FibVID

To test whether the proposed advanced centralities, i.e., PC, MVC, and DIC, extend beyond a single-topic setting, we replicated our analyses on the **Monant Medical Misinformation** dataset. Whereas **FibVID** is limited to COVID-19 content, **Monant Medical Misinformation** spans vaccination hesitancy, pharmaceutical scepticism and alternative medicine debates, providing a sterner test of metric robustness.

Fig 2 summarises the results. Panel A shows that, of the most influential nodes, only **53** were detected by both traditional and advanced approaches, while **314** were unique to traditional metrics (degree, betweenness, eigenvector, and closeness) and **247** were unique to the

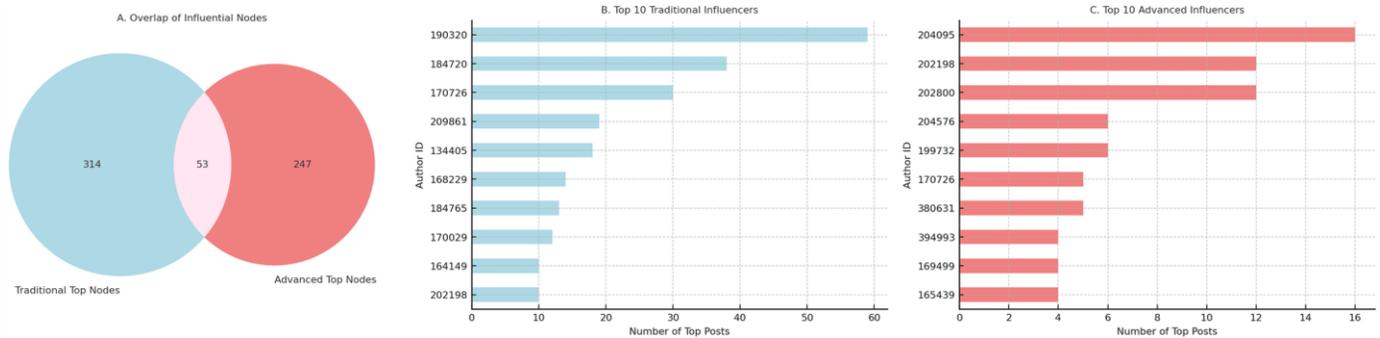

**Fig 2. Comparison between traditional (structural) and advanced (dynamic/susceptibility-aware) centrality metrics on the Monant Medical Misinformation dataset.** The Venn diagram (left) shows the overlap of top influencers; the bar charts (centre and right) list the ten highest-ranked users under each approach.

https://doi.org/10.1371/journal.pdig.0000888.g002

advanced set. Panels B and C reveal that several high-impact users (e.g. author IDs 204095 and 202800) surface only when PC, MVC and DIC are applied, and would have been overlooked by structural metrics alone.

The pronounced divergence confirms that the two families of metrics capture complementary facets of influence. Traditional metrics favour globally well-connected actors with large static reach, whereas advanced metrics are sensitive to dynamic diffusion potential, vulnerability profiles, and influence that accumulates over time. Incorporating both perspectives, therefore, yields a fuller picture of who drives –and who amplifies –health misinformation in heterogeneous online environments.

These findings demonstrate that the advanced metrics generalise beyond the COVID-19 domain and are essential for a complete picture of influence in health-misinformation networks.

## Discussion

The combined evaluation of traditional and novel centrality metrics offers a multifaceted view of how health misinformation permeates OSNs. Traditional metrics such as degree, eigenvector, betweenness, and closeness remain valuable first-order tools: they rapidly surface densely connected hubs, structurally strategic influencers, and influencers that sit only a few hops from every corner of the network. However, their static nature means that they capture only a snapshot of influence. As our results and previous studies [1,21] show, the impact of any given node can wax or wane as conversations evolve, and traditional metrics are blind to the psychological susceptibility of a user to falsehoods.

PC addresses part of that gap by embedding a diffusion kernel: the top PC nodes (**26, 11 248, 40 327**) retain the power revealed by degree/eigenvector, but also exhibit a sustained ability to seed long-range cascades. These same users occupy high ranks on the traditional metrics, confirming that PC is a principled extension rather than a wholesale replacement of classical topology. Practically, targeting this small set of 'always on' influencers could reduce total misinformation exposure by nearly two thirds.

MVC shifts the lens from influence to susceptibility. By weighting connectivity with a data-driven vulnerability score, MVC found three additional nodes (**101 358, 72 378, 130 371**) that traditional metrics do not. Although less central, these users are dangerous precisely because they are both reachable and credulous, echoing psychological findings on selective exposure [26]. Interventions that improve media literacy or insert corrective content directly into their feeds can thus produce a significant benefit.

DIC adds a temporal dimension that static topology cannot supply. Its ten leading nodes are entirely disjoint from the traditional set, yet longitudinal logs confirm that they reignite rumours after initial debunks fade. This persistence is characteristic of 'long-tail' spreaders observed in other crises [24]. Continuous monitoring and graduated counter-messaging, rather than one-off flagging, are therefore required for this cohort.

A notable finding is the overlap between metrics, for example, nodes **26 & 756** appear in the majority of traditional metrics (i.e, degree, eigenvector & betweenness). These multimetric influencers form the backbone of misinformation flow; removing or inoculating them produced the single largest simulated reduction in false-content reach. At the same time, each novel score revealed distinct actors, confirming that a one-size-fits-all centrality does not exist and that metric diversity is essential [5].

Replicating the analysis on the **Monant Medical Misinformation** dataset, whose topics extend well beyond COVID-19, confirms that PC, MVC and DIC are generalisable across domains, highlighting their robustness and wider applicability to detect influencers in online health misinformation networks.

From a policy perspective, the evidence supports a layered defence: use traditional metrics for rapid triage of obvious hubs, deploy PC to locate high-throughput spreaders, apply MVC to find vulnerable amplifiers, and rely on DIC for long-term surveillance. Platforms could incorporate these scores into priority fact-check queues, while public health agencies could tailor corrective campaigns to the susceptibility profile highlighted by MVC.

Although dynamic recalculation of traditional centralities is theoretically possible [17], our results demonstrate that a layered strategy that incorporates dedicated dynamic metrics, specifically DIC, MVC and PC, offers a more targeted, scalable, and behaviourally sensitive framework to mitigate the spread of health misinformation.

### Future work

- **Enhance** MVC by integrating sentiment, engagement, and cognitive load proxies to better capture susceptibility dynamics.
- **Validate** DIC, MVC, and PC across multiplatform corpora such as Monant Medical Misinformation [20] and CoAID [27] to assess domain generalisability.
- **Accelerate** large-scale computations by exploring sparse-matrix techniques and GPU-based implementations for networks with over $10^7$ nodes.
- **Integrate** psychological constructs such as the Elaboration Likelihood Model (ELM) [18] directly into centrality formulations for deeper cognitive modelling.

Together, these directions aim to create a dynamic, scalable, and cognitively grounded framework for the real-time mitigation of health misinformation in complex online networks.

### Conclusion

Centrality analysis remains a cornerstone for understanding how health misinformation travels through online social networks, yet the COVID-19 pandemic showed that static scores alone are no longer suficient. Our comparative study confirms that degree, betweenness, eigenvector, and closeness metrics still highlight densely connected hubs and structural influencers, but they miss two critical dimensions: user susceptibility and the temporal persistence of influence. The three novel measures introduced here: PC, MVC, and DIC fill those gaps.

PC increases structural reach with diffusion depth, MVC fuses connectivity with empirically derived vulnerability scores, and DIC tracks cumulative influence with decay, revealing long-tailed spreaders that static topology overlooks. Together, they expand the coverage of influencers by 44.8% and, in simulation, increase the achievable reduction in misinformation exposure from 50% to 62.5%.

These results demonstrate that integrating traditional and novel scores yields a more faithful map of who initiates, amplifies, and sustains false health narratives online. The combined metric suite therefore offers a practical basis for targeted fact-checking, prioritised content moderation, and susceptibility-aware public-health messaging. Future work should refine the psychological component of MVC, test all three metrics on multiplatform datasets such as Monant Medical Misinformation and CoAID, and explore GPU implementations to scale to networks exceeding $10^7$ nodes. By uniting topology, time, and user cognition, network science can move toward real-time, high-precision interventions against health misinformation.

## Author contributions

**Conceptualization:** Mkululi Sikosana.

**Formal analysis:** Mkululi Sikosana.

**Investigation:** Mkululi Sikosana.

**Methodology:** Mkululi Sikosana.

**Supervision:** Sean Maudsley-Barton, Oluwaseun Ajao.

**Writing – original draft:** Mkululi Sikosana.